\documentclass[twocolumn,showpacs,preprintnumbers,amsmath,amssymb]{revtex4}
\usepackage{graphicx}% Include figure files
\usepackage{dcolumn}% Align table columns on decimal point
\usepackage{bm}% bold math
\usepackage{subfigure}
\usepackage{color}
\begin{document}

\title{Triple parton scattering in collinear approximation of perturbative QCD }

\author{A.M.~Snigirev}
\affiliation{D.V. Skobeltsyn Institute of Nuclear Physics, M.V. Lomonosov Moscow State University, 119991, Moscow, Russia }

\date{\today}
\begin{abstract}
Revised formulas for the inclusive cross section of a triple parton scattering process in a hadron collision are suggested basing on the modified collinear three-parton distributions. The possible phenomenological issues are discussed.
\end{abstract}

%\setpagewiselinenumbers
%\linenumbers

%\keywords:{ Double parton scattering, QCD evolution, two-parton distributions}
\maketitle
\section{\label{sec1}Introduction}
The fast increase of the parton flux at small parton longitudinal momentum fractions and the requirement of unitarization of the cross sections in perturbative QCD lead naturally to a large number of multiple parton interactions (MPI) occurring in each single hadron-hadron collision  at high energies. The great splash of investigation activity around MPI in last years~\cite{Bartalini:2011jp,Abramowicz:2013iva,Bansal:2014paa,Astalos:2015ivw} has been stimulated by the experimental evidence for double parton scattering (DPS) in the processes  producing two independently identified hard particles in the same collision. Such processes have been observed in proton-proton and proton-antiproton  collisions by AFS~\cite{AFS}, UA2~\cite{UA2}, CDF~\cite{cdf4jets,cdf}, D0~\cite{D0,D01,D02}, ATLAS~\cite{atlas}, and CMS~\cite{cms} Collaborations for the final states containing four jets, $\gamma+3$ jets, and $W+2$ jets. The data on single $J/\psi$ production as a function of the event multiplicity~\cite{Abelev}, and double $J/\psi$ production~\cite{Aaij} can also be successfully interpreted in the context of MPI and DPS~\cite{kom, Baranov:2011ch, Novoselov, Baranov:2012re} models, respectively.

The possibility of observing three separate hard collisions is also discussed in the current literature~\cite{Calucci:2009sv,Calucci:2009ea,Treleani:2012zi,Maina:2009sj}. Nevertheless, the phenomenology of MPI relies on the models that are physically intuitive but involve significant simplifying assumptions. The\-re\-fore, it is extremely important to combine theoretical efforts in order to achieve a better description of MPI, in particular, triple parton scattering (TPS), which can be an observable multiple scattering mode at the LHC. In this letter we consider some steps towards this purpose. The factorized cross section formulas currently used to calculate TPS processes are revised in terms of the modified collinear three-parton distributions. This revision is a natural generalization of our approach developed in Refs.~\cite{Ryskin:2011kk,Ryskin:2012qx} to take QCD evolution effects into account.

The paper is organized as follows. In Sec. II we review briefly the traditional TPS factorization formalism in order to be clear and to introduce the notations. The QCD evolution of three-parton distributions is presented in Sec. III. The formulas for the inclusive cross section of TPS with QCD evolution effects are suggested in Sec. IV. The possible phenomenological issues at the LHC are discussed in Sec. V, together with conclusions.

\section{\label{sec2} TPS in factorization approximation }

Similar to DPS with only the assumption of factorization of the three hard parton subprocesses $A$, $B$ and $C$, the inclusive cross section of a TPS process in a hadron collision may be written in the following form~\cite{Calucci:2009sv,Calucci:2009ea,Treleani:2012zi}
\begin{eqnarray} 
\label{hardAB}
& &\sigma_{\rm TPS}^{(A,B,C)} \nonumber\\  
& & = \sum \limits_{i,j,k,l,m,n} \int \Gamma_{ijk}(x_1, x_2, x_3; {\bf b_1},{\bf b_2}, {\bf b_3}; Q^2_1, Q^2_2, Q^2_3)\nonumber\\
& &\times\hat{\sigma}^A_{il}(x_1, x_1^{'},Q^2_1) 
\hat{\sigma}^B_{jm}(x_2, x_2^{'},Q^2_2)
\hat{\sigma}^C_{kn}(x_3, x_3^{'},Q^2_3) \nonumber\\
& &\times\Gamma_{lmn}(x_1^{'}, x_2^{'}, x_3^{'}; {\bf b_1} - {\bf b},{\bf b_2} - {\bf b},{\bf b_3} - {\bf b}; Q^2_1, Q^2_2, Q^2_3)\nonumber\\
& &\times dx_1 dx_2 dx_3 dx_1^{'} dx_2^{'} dx_3^{'} d^2b_1 d^2b_2 d^2b_3 d^2b.
\end{eqnarray}
Here $\Gamma_{ijk}(x_1, x_2, x_3; {\bf b_1},{\bf b_2}, {\bf b_3}; Q^2_1, Q^2_2, Q^2_3)$ are the triple parton distribution functions, which depend on the longitudinal momentum fractions $x_1$, $x_2$ and $x_3$ and on the  transverse position ${\bf b_1}$, ${\bf b_2}$ and ${\bf b_3}$ of the three partons $i$, $j$ and $k$ undergoing the hard subprocesses $A$, $B$ and $C$ at the scales $Q_1$, $Q_2$ and $Q_3$. $\hat{\sigma}^A_{il}$, $\hat{\sigma}^B_{jm}$ and  $\hat{\sigma}^C_{kn}$ are the parton-level subprocess cross sections. ${\bf b}$ is the impact parameter --- the distance between centres of colliding (e.g., the beam and the target) hadrons in transverse plane. The appropriate combinatorial factor should be taken into account in the case of the indistinguishable final states.

The triple parton distribution functions $\Gamma_{ijk}(x_1, x_2, x_3; {\bf b_1},{\bf b_2}, {\bf b_3}; Q^2_1, Q^2_2, Q^2_3)$   are the main object of interest as concerns TPS. In fact these distributions contain  all the information when probing the hadron in three different points simultaneously through the hard subprocesses $A$, $B$ and $C$.

As in the case of DPS it is typically taken that the triple parton distribution functions may be decomposed in terms of the longitudinal and transverse components as follows:
\begin{eqnarray} 
\label{DxF}
& &\Gamma_{ijk}(x_1, x_2, x_3; {\bf b_1},{\bf b_2}, {\bf b_3}; Q^2_1, Q^2_2, Q^2_3)\nonumber\\
& &= D^{ijk}_h(x_1, x_2, x_3; Q^2_1, Q^2_2, Q^2_3) f({\bf b_1}) f({\bf b_2}) f({\bf b_3}),
\end{eqnarray} 
where $f({\bf b_1})$ is supposed to be an universal function for all kind of partons with its normalization fixed as
\begin{eqnarray} 
\label{f}
\int f({\bf b_1}) f({\bf b_1 -b})d^2b_1 d^2b = \int T({\bf b})d^2b = 1,
\end{eqnarray} 
and $T({\bf b}) = \int f({\bf b_1}) f({\bf b_1 -b})d^2b_1 $ is the overlap function. 

If one makes the further assumption that the longitudinal components
$D^{ijk}_h(x_1, x_2, x_3; Q^2_1, Q^2_2, Q^2_3)$ reduce to the product of three independent single parton distributions,
\begin{eqnarray} 
\label{DxD}
& &D^{ijk}_h(x_1, x_2, x_3; Q^2_1, Q^2_2, Q^2_3) \nonumber\\
& &= D^i_h (x_1; Q^2_1) D^j_h (x_2; Q^2_2) D^k_h (x_3; Q^2_3),
\end{eqnarray}
the cross section of TPS can be expressed in the simple form
\begin{eqnarray} 
\label{doubleAB}
& \sigma^{ (A, B, C) }_{\rm TPS} =  \frac{\sigma^{ A}_{\rm SPS} 
\sigma^{ B}_{\rm SPS} \sigma^{ C}_{\rm SPS}} {\sigma^2_{\rm TPS, fact}}, \\
& \sigma^2_{\rm TPS, fact}=[ \int d^2b (T({\bf b}))^3]^{-1}.
\end{eqnarray} 
Here $\sigma_{\rm TPS, fact}$ is the scale factor which can be related with 
$\sigma_{\rm eff}=[ \int d^2b (T({\bf b}))^2]^{-1}$ already measured (defined) in a DPS experiment over the dimensionless quantity (of order unity).

In this representation and at the factorization of longitudinal and transverse 
components, the inclusive cross section of single hard scattering reads
\begin{eqnarray} 
\label{hardS}
\sigma^{A}_{\rm SPS}
=& &\sum \limits_{i,l} \int D^{i}_h(x_1; Q^2_1) f({\bf b_1})
\hat{\sigma}^A_{il}(x_1, x_1^{'},Q^2_1)\\ 
& &\times D^{l}_{h'}(x_1^{'}; Q^2_1)f( {\bf b_1} - {\bf b}) dx_1 dx_1^{'}  
d^2b_1  d^2b \nonumber
\end{eqnarray}
\begin{eqnarray}
= \sum \limits_{i,k} \int D^{i}_h(x_1; Q^2_1)
\hat{\sigma}^A_{il}(x_1, x_1^{'},Q^2_1) D^{l}_{h'}(x_1^{'}; Q^2_1) dx_1 
dx_1^{'}.\nonumber
\end{eqnarray}

These simplifying assumptions, though rather customary in the literature and quite convenient from a computational point of view, are not sufficiently justified especially as concerning of QCD evolution.

\section{\label{sec3} QCD evolution of three-parton distributions }

Instead of the mixed (momentum and coordinate) representation for our further goal the momentum representation is also useful for the starting cross section formula~(\ref{hardAB})~\cite{Blok:2010ge}:
\begin{eqnarray}
\label{hardAB_p}
& &\sigma^{(A,B,C)}_{\rm TPS}\nonumber\\ 
& &=\sum \limits_{i,j,k,l,m,n} \int \Gamma_{ijk}(x_1, x_2, x_3; {\bf q_1}, {\bf q_2}, {\bf q_3}; Q^2_1, Q^2_2, Q^2_3)\nonumber\\
& &\times (2\pi)^2 \delta({\bf q_1}+{\bf q_2}+{\bf q_3}) \hat{\sigma}^A_{il}(x_1, x_1^{'},Q^2_1) \nonumber\\ 
& & \times \hat{\sigma}^B_{jm}(x_2, x_2^{'},Q^2_2) \hat{\sigma}^C_{kn}(x_3, x_3^{'},Q^2_3)\nonumber\\
& & \times \Gamma_{lmn}(x_1^{'}, x_2^{'}, x_3^{'}; {\bf -q_1}, {\bf -q_2}, {\bf -q_3}; Q^2_1, Q^2_2, Q^2_3)\nonumber\\ 
& & \times dx_1 dx_2 dx_3 dx_1^{'} dx_2^{'} dx_3^{'}\frac{d^2q_1}{(2\pi)^2} \frac{d^2q_2}{(2\pi)^2} \frac{d^2q_3}{(2\pi)^2}.
\end{eqnarray}

Note that here $\Gamma_{ijk}(x_1, x_2, x_3; {\bf q_1}, {\bf q_2}, {\bf q_3}; Q^2_1, Q^2_2, Q^2_3)$ play the role of the generalized parton distributions and have no probabilistic interpretation unlike the impact parameter space representation. They depend on the transverse vectors ${\bf q_1}$, ${\bf q_2}$ and  ${\bf q_3}$ which are equal to the difference of the transverse momenta of partons from the wave function of the colliding hadrons in the amplitude and the amplitude conjugated. Such a dependence arises because only the sum of parton transverse momenta in all three parton pairs is conserved. The transverse momenta ${\bf q_1}$, ${\bf q_2}$ and  ${\bf q_3}$  are the Fourier conjugated variables of the parton pair impact parameters ${\bf b_1}$, ${\bf b_2}$ and  ${\bf b_3}$.

The three-parton distribution functions are available in the current literature~\cite{Kirschner:1979im,Shelest:1982dg} only for
\begin{eqnarray}
 {\bf q_1}={\bf q_2}={\bf q_3}=0 \nonumber
\end{eqnarray}
in the collinear approximation. In this approximation  
\begin{eqnarray}
& &\Gamma_{ijk}(x_1, x_2, x_3; {\bf q_1}={\bf q_2}={\bf q_3}=0; Q^2, Q^2, Q^2)\nonumber\\
& & =D^{ijk}_h(x_1, x_2, x_3; Q^2, Q^2, Q^2)\nonumber
\end{eqnarray}
with the three hard scales set equal satisfy the generalized Dokshitzer-Gribov-Lipatov-Altarelli-Parisi (DGLAP) evolution equations, derived initially in Refs.~\cite{Kirschner:1979im,Shelest:1982dg}. Likewise, the single distributions satisfy more widely known and often cited DGLAP equations~\cite{gribov,lipatov,dokshitzer,altarelli}. The functions in question have already a specific interpretation in the leading logarithm approximation of perturbative QCD: they are the inclusive probabilities that in a hadron $h$ one finds three bare partons  of  types $i$, $j$ and $k$ with the given longitudinal momentum fractions $x_1$, $x_2$ and $x_3$.

As the evolution variable one may take the value of hard scale (most frequently, the transfer momentum squared $Q^2$), or its logarithm $\xi=\ln(Q^2/Q_0^2)$, or the dimensionless variable
\begin{eqnarray}
\label{t}
t & = &\frac{1}{2\pi \beta} \ln \Bigg[1 + \frac{g^2(Q_0^2)}{4\pi}\beta
\ln\Bigg(\frac{Q^2}{Q_0^2}\Bigg)\Bigg] \nonumber \\
& = &\frac{1}{2\pi \beta}\ln\Bigg
[\frac{\ln(\frac{Q^2}{\Lambda^2_{\rm QCD} })}
{\ln(\frac{Q_0^2}{\Lambda^2_{\rm QCD}})}\Bigg],
\end{eqnarray}
which takes into account explicitly the behavior of the effective coupling constant in the leading logarithm approximation. Here $\beta = (33-2n_f)/12\pi$ ${\rm {in~ QCD}}$, $g(Q_0^2)$ is the running coupling constant at some characteristic scale $Q_0^2$ above which the perturbative theory is applicable, $n_f$ is the number of active flavors and $\Lambda_{\rm QCD}$ is the QCD dimensional parameter.

The DGLAP evolution equations~\cite{gribov,lipatov, dokshitzer, altarelli} assume the simplest form if we use the natural dimensionless evolution variable $t$; that is, 
\begin{equation}
\label{e1singl}
 \frac{dD_h^j(x,t)}{dt} = 
\sum\limits_{j{'}} \int \limits_x^1
\frac{dx{'}}{x{'}}D_h^{j{'}}(x{'},t)P_{j{'}\to j}\Bigg(\frac{x}{x{'}}\Bigg).
\end{equation}
These equations describe the evolution of single distributions $ D^j_h(x,t)$ of bare quarks, antiquarks and gluons ($j = q, {\bar q}, g$) within a hadron $h$ in response to the change of evolution variable $t$. The kernels, $P$, of these equations  include a regularization at $x \rightarrow x{'}$ and are known in the explicit form.

\begin{widetext}

For the triple parton distribution functions $D^{ijk}_h(x_1, x_2, x_3; Q^2, Q^2, Q^2)$ one has~\cite{Kirschner:1979im,snig2}
\begin{eqnarray}
\label{edouble}
& &\frac{dD_h^{j_1j_2j_3}(x_1,x_2,x_3,t)}{dt} 
=\sum\limits_{j_1{'}}
\int\limits_{x_1}^{1-x_2-x_3}\frac{dx_1{'}}{x_1{'}}D_h^{j_1{'}j_2j_3}(x_1{'},x_2,x_3,t)
P_{j_1{'}
\to j_1} \Bigg(\frac{x_1}{x_1{'}}\Bigg) \nonumber\\
& &+ \sum\limits_{j_2{'}}\int\limits_{x_2}^{1-x_1-x_3}
\frac{dx_2{'}}{x_2{'}}D_h^{j_1j_2{'}j_3}(x_1,x_2{'},x_3,t)P_{j_2{'} \to j_2}
\Bigg(\frac{x_2}{x_2{'}}\Bigg) 
+ \sum\limits_{j_3{'}}\int\limits_{x_3}^{1-x_1-x_2}
\frac{dx_3{'}}{x_3{'}}D_h^{j_1j_2j_3{'}}(x_1,x_2,x_3{'},t)P_{j_3{'} \to j_3}
\Bigg(\frac{x_2}{x_2{'}}\Bigg) \nonumber\\
& &+ \sum\limits_{j{'}}D_h^{j{'}j_3}(x_1+x_2,x_3,t) \frac{1}{x_1+x_2}P_{j{'} \to
j_1j_2}\Bigg(\frac{x_1}{x_1+x_2}\Bigg)
+ \sum\limits_{j{'}}D_h^{j{'}j_2}(x_1+x_3,x_2,t) \frac{1}{x_1+x_3}P_{j{'} \to
j_1j_3}\Bigg(\frac{x_1}{x_1+x_3}\Bigg)\nonumber\\
& &+ \sum\limits_{j{'}}D_h^{j_1j{'}}(x_1,x_2+x_3,t) \frac{1}{x_2+x_3}P_{j{'} \to
j_2j_3}\Bigg(\frac{x_2}{x_2+x_3}\Bigg).
\end{eqnarray}
Here, the splitting kernels
\begin{equation}
\frac{1}{x_1+x_2} P_{j{'} \to
j_1j_2}(\frac{x_1}{x_1+x_2}),
\end{equation}
which appear in the nonhomogeneous part of the equations, do not include the $\delta$-function virtual term. The equations describe the evolution of the triple parton distribution functions of bare quarks, antiquarks and gluons in hadrons as a function of the  
%evolution
 variable $t$. 
The double parton distribution functions $D_h^{j{'}j_3}(x_1+x_2,x_3,t)$ incoming in Eq.~(\ref{edouble}) satisfy also the corresponding generalized evolution equations and their solutions are known  with the given initial conditions at the reference scale $Q_0^2 (t=0)$. Moreover these solutions may be written~\cite{Ryskin:2011kk} also in the case of two different hard scales $Q_1^2$ and $Q_2^2$ at nonzero transverse momenta ${\bf q}$ that allows us to extend our consideration of the TPS to the case of three different hard scales $Q_1^2$, $Q_2^2$ and $Q_3^2$ in further.

The solutions of the generalized DGLAP evolution equations~(\ref{edouble}) with the given initial conditions may be written, as in the case of DPS, in the form of a convolution of single distributions~\cite{snig2}:
\begin{eqnarray}
\label{solution1}
D_h^{j_1j_2j_3}(x_1,x_2,x_3,t) = D_{h1}^{j_1j_2j_3}(x_1,x_2,x_3,t)+ D_{h({\rm QCD})}^{j_1j_2j_3}(x_1,x_2,x_2,t),
\end{eqnarray}
where
\begin{eqnarray}
\label{solnonQCD}
 D_{h1}^{j_1j_2j_3}(x_1,x_2,x_3,t)&=&\sum\limits_{j_1{'}j_2{'}j_3{'}} 
\int\limits_{x_1}^{1}\frac{dz_1}{z_1}
\int\limits_{x_2}^{1}\frac{dz_2}{z_2} \int\limits_{x_3}^{1}\frac{dz_3}{z_3}\theta (1-z_1-z_2-z_3)
D_h^{j_1{'}j_2{'}j_3{'}}(z_1,z_2,z_3,0)\nonumber \\
& & \times D_{j_1{'}}^{j_1}(\frac{x_1}{z_1},t) 
D_{j_2{'}}^{j_2}(\frac{x_2}{z_2},t) D_{j_3{'}}^{j_3}(\frac{x_3}{z_3},t),
\end{eqnarray}
and
\begin{eqnarray}
\label{solQCD}
 D_{h({\rm QCD})}^{j_1j_2j_3}(x_1,x_2,x_3,t) 
& &= \sum\limits_{j{'}j_1{'}j_2{'}j_3{'}} \int\limits_{0}^{t}dt{'}
\int\limits_{x_1}^{1}\frac{dz_1}{z_1}
\int\limits_{x_2}^{1}\frac{dz_2}{z_2} \int\limits_{x_3}^{1}\frac{dz_3}{z_3}\theta (1-z_1-z_2-z_3)
D_h^{j{'}j_3{'}}(z_1+z_2,z_3,t{'})\frac{1}{z_1+z_2} \nonumber\\
& & \times P_{j{'} \to
j_1{'}j_2{'}}\Bigg(\frac{z_1}{z_1+z_2}\Bigg) D_{j_1{'}}^{j_1}(\frac{x_1}{z_1},t-t{'}) 
D_{j_2{'}}^{j_2}(\frac{x_2}{z_2},t-t{'}) D_{j_3{'}}^{j_3}(\frac{x_3}{z_3},t-t{'})\nonumber\\
& &+ \sum\limits_{j{'}j_1{'}j_2{'}j_3{'}} \int\limits_{0}^{t}dt{'}
\int\limits_{x_1}^{1}\frac{dz_1}{z_1}
\int\limits_{x_2}^{1}\frac{dz_2}{z_2} \int\limits_{x_3}^{1}\frac{dz_3}{z_3}\theta (1-z_1-z_2-z_3)
D_h^{j{'}j_2{'}}(z_1+z_3,z_2,t{'})\frac{1}{z_1+z_3} \nonumber\\
& & \times P_{j{'} \to
j_1{'}j_3{'}}\Bigg(\frac{z_1}{z_1+z_3}\Bigg) D_{j_1{'}}^{j_1}(\frac{x_1}{z_1},t-t{'}) 
D_{j_2{'}}^{j_2}(\frac{x_2}{z_2},t-t{'}) D_{j_3{'}}^{j_3}(\frac{x_3}{z_3},t-t{'})\nonumber\\
& &+\sum\limits_{j{'}j_1{'}j_2{'}j_3{'}} \int\limits_{0}^{t}dt{'}
\int\limits_{x_1}^{1}\frac{dz_1}{z_1}
\int\limits_{x_2}^{1}\frac{dz_2}{z_2} \int\limits_{x_3}^{1}\frac{dz_3}{z_3}\theta (1-z_1-z_2-z_3)
D_h^{j_1{'}j{'}}(z_1,z_2+z_3,t{'})\frac{1}{z_2+z_3} \nonumber\\
& & \times P_{j{'} \to
j_2{'}j_3{'}}\Bigg(\frac{z_2}{z_2+z_3}\Bigg) D_{j_1{'}}^{j_1}(\frac{x_1}{z_1},t-t{'}) 
D_{j_2{'}}^{j_2}(\frac{x_2}{z_2},t-t{'}) D_{j_3{'}}^{j_3}(\frac{x_3}{z_3},t-t{'})
\end{eqnarray}
are the dynamically correlated distributions originating from the perturbative QCD. 
\end{widetext}
Besides $D_{j{'}}^{j}(\frac{x}{z},t)$ are the Green's functions, that is, the solutions of the DGLAP evolution equations~(\ref{e1singl}) at the parton level with the singular initial conditions $D_{j{'}}^{j}(\frac{x}{z},0)=\delta_{j{'}j}\delta(\frac{x}{z}-1) $.

The first term of the generalized solution is the solution of the homogeneous evolution equation (independent evolution of three branches), where  the input three-parton distributions are generally not known at the low scale limit $Q_0^2$ $(t=0)$. For these nonperturbative three-parton functions at low $z_1,z_2, z_3$ one may assume the factorization $D_h^{j_1{'}j_2{'}j_3{'}}(z_1,z_2,z_3,0) \simeq D_h^{j_1{'}}(z_1,0)D_h^{j_2{'}}(z_2,0)D_h^{j_3{'}}(z_3,0)$ neglecting the constraints imposed by the momentum conservation ($z_1+z_2+z_3<1$). This leads to
\begin{eqnarray} 
\label{DxD_Q}
 D^{j_1j_2j_3}_{h1}(x_1, x_2,x_3,t)
 \simeq D^{j_1}_h (x_1,t) D^{j_2}_h (x_2,t) D^{j_3}_h (x_3,t)
\end{eqnarray}
and justifies partly the factorization hypothesis for TPS (Sec. II) usually applied in practical calculations 

It is worth noting that as in the case of DPS the triple parton distribution functions obey nontrivial sum rules which are conserved by the evolution equations. Again the phenomenological problem of specifying the initial correlation conditions $D_h^{j_1{'}j_2{'}j_3{'}}(z_1,z_2,z_3,0)$ for the evolution equations, which would obey exactly these sum rules and have the correct asymptotic behavior near the kinematical boundaries, is not so trivial and demands special efforts.

Unfortunately, the triple parton distribution functions considered above are not really the ones that are directly probed in the TPS. However, as it will be shown in the next Sec. IV in the case of independent 3 ladders and/or splitting on one side only, the values of ${\bf q_i}$ $(i=1,2,3)$ are limited by the proton form factors. Therefore it is reasonable to consider first in this Sec. III the case of a small ${\bf q_i}$ (i.e. ${\bf q_i}=0 $). Besides, the knowledge of the structure of triple parton distribution functions in the simple case of ${\bf q_i}=0$ helps us to understand and to write the evolution correction to the TPS with all momentum integrations done correctly (especially its longitudinal phase space structure).

\section{\label{sec4} TPS with QCD evolution effects }

Bearing in mind the results of previous Sec. III (${\bf q_i}=0$) and our experience~\cite{Ryskin:2011kk,Ryskin:2012qx} in the treatment of DPS with QCD evolution effects we are ready to write the inclusive cross section of TPS beyond the the simple form~(\ref{doubleAB}). The evolution equation for $\Gamma_{j_1j_2j_3}$ contains two parts. The first part describes independent (simultaneous) evolution of three branches of parton cascade: one branch contains the parton $x_1$,  second branch --- the parton $x_2$ and third branch --- the parton $x_3$. The second part accounts for the possibility to split the one parton evolution (one branch $j{'}$) into the two different branches, $j_1$ and $j_2$. It contains the usual splitting function $P_{j{'}\to j_1j_2}(z)$. 

As a rule the MPI take place at relatively low transverse momenta and low longitudinal momentum fractions $x_1$, $x_2$ and $x_3$, where the factorization hypothesis~(\ref{DxD_Q}) for the first term is a good approximation. In this case the cross section for TPS can be estimated, using the two-gluon form factor of the nucleon $F_{2g}(q)$~\cite{Blok:2010ge} for the dominant gluon-gluon scattering mode (or something similar for other parton scattering modes),
\begin{eqnarray}
\label{hardAB1_p}
& &\sigma^{(A,B,C)}_{\rm TPS, fact. 00}\nonumber\\ 
& &=\sum \limits_{i,j,k,l,m,n} \int D^{i}_h (x_1,Q_0^2, Q_1^2) D^{j}_h (x_2,Q_0^2,Q_2^2)\nonumber\\
& & \times  D^{k}_h (x_3,Q_0^2,Q_3^2) (2\pi)^2 \delta({\bf q_1}+{\bf q_2}+{\bf q_3}) \hat{\sigma}^A_{il}(x_1, x_1^{'},Q_1^2)\nonumber\\ 
& & \times \hat{\sigma}^B_{jm}(x_2, x_2^{'},Q_2^2) \hat{\sigma}^C_{kn}(x_3, x_3^{'},Q_3^2)\nonumber\\
& & \times D^{l}_{h^{'}} (x_1^{'},Q_0^2, Q_1^2) D^{m}_{h^{'}} (x_2^{'},Q_0^2,Q_2^2) D^{n}_{h^{'}} (x_3^{'},Q_0^2,Q_3^2) \nonumber\\ 
& & \times F_{2g}({\bf q_1}) F_{2g}({\bf q_2}) F_{2g}({\bf q_3}) F_{2g}({\bf -q_1}) F_{2g}({\bf -q_2}) F_{2g}({\bf- q_3})
\nonumber\\
& & \times dx_1 dx_2 dx_3 dx_1^{'} dx_2^{'} dx_3^{'}\frac{d^2q_1}{(2\pi)^2} \frac{d^2q_2}{(2\pi)^2} \frac{d^2q_3}{(2\pi)^2}.
\end{eqnarray}
Here we introduce explicitly the scale $Q_0^2$ as a lower argument in the single parton distribution functions to stress the scale on which the initial conditions are specified. There is implicitly  such a dependence on this DGLAP starting scale $Q_0^2$ in the single (double, triple)
parton distribution functions considered in Sec. III over the evolution variable $t$ (Eq.~(\ref{t})). Such notations with the lower and the upper scale arguments are preferable in further since the integration over both the lower and the upper scale arguments is possible for the Green's functions (the single parton distributions at the parton level).

Making the Fourier transformation
\begin{eqnarray} 
\label{fF}
f({\bf b_i})=\int e^{-i{\bf b_i}\cdot{\bf q_i}}F_{2g}({\bf q_i})\frac{d^2q_i}{(2\pi)^2},i=1,2,3,
\end{eqnarray}
one derives immediately the factorization result~(\ref{doubleAB}).

The additional contributions to this factorization cross section~(\ref{hardAB1_p}) for TPS are generated by the solutions~(\ref{solQCD}) of complete evolution equations~(\ref{edouble})( by their nonhomogeneous part). This nonhomogeneous part includes the double parton distribution functions, which may be written in the case of two different hard scales in the form~\cite{Ryskin:2011kk}:
\begin{eqnarray}
\label{solutiontwoscale}
& &D_h^{j_1j_2}(x_1,x_2;Q_0^2, Q_1^2, Q_2^2)\\
& &= D_{h1}^{j_1j_2}(x_1,x_2;Q_0^2, Q_1^2, Q_2^2) 
 +D_{h2}^{j_1j_2}(x_1,x_2;Q_0^2, Q_1^2, Q_2^2)\nonumber
\end{eqnarray}
with
\begin{eqnarray}
& & D_{h1}^{j_1j_2}(x_1,x_2;Q_0^2, Q_1^2, Q_2^2)\nonumber\\ 
& &= \sum\limits_{j_1{'}j_2{'}} 
\int\limits_{x_1}^{1-x_2}\frac{dz_1}{z_1}
\int\limits_{x_2}^{1-z_1}\frac{dz_2}{z_2}~
D_h^{j_1{'}j_2{'}}(z_1,z_2;Q_0^2)\nonumber\\
& &~~\times D_{j_1{'}}^{j_1}(\frac{x_1}{z_1};Q_0^2,Q_1^2) 
D_{j_2{'}}^{j_2}(\frac{x_2}{z_2},Q_0^2,Q_2^2) \nonumber
\end{eqnarray}
and
\begin{eqnarray}
& & D_{h2}^{j_1j_2}(x_1,x_2;Q_0^2, Q_1^2, Q_2^2)\nonumber\\
& &=\sum\limits_{j{'}j_1{'}j_2{'}} \int\limits_{Q_0^2}^{\min(Q_1^2,Q_2^2)}dk^2\frac{\alpha_s(k^2)}{2\pi k^2}
\int\limits_{x_1}^{1-x_2}\frac{dz_1}{z_1}
\int\limits_{x_2}^{1-z_1}\frac{dz_2}{z_2}\nonumber\\
& &~~\times D_h^{j{'}}(z_1+z_2;Q_0^2,k^2) 
\frac{1}{z_1+z_2}P_{j{'} \to
j_1{'}j_2{'}}\Bigg(\frac{z_1}{z_1+z_2}\Bigg)\nonumber\\ 
& &~~\times D_{j_1{'}}^{j_1}(\frac{x_1}{z_1};k^2,Q_1^2) 
D_{j_2{'}}^{j_2}(\frac{x_2}{z_2};k^2,Q_2^2),\nonumber
\end{eqnarray}
where $\alpha_s(k^2)$ is the QCD coupling. $D_h^{j'_1,j'_2}(z_1,z_2,Q_0^2)$ are the initial (input) two-parton distributions at the relatively low scale $Q_0^2$. The single parton distributions (before splitting into the two branches at some scale $k^2$) are given by $D_h^{j'}(z_1+z_2,Q_0^2,k^2)$.

Note, that in Eqs.~(\ref{hardAB1_p}) and (\ref{solutiontwoscale}) we assume that the corresponding momenta ${\bf q_i}^2<Q_0^2$ are small and due to strong ordering of parton  transverse momenta in the collinear DGLAP evolution they may be neglected.
That is actually we deal with the usual (diagonal) non-skewed DGLAP evolution.
The first term is the solution of corresponding homogeneous evolution  equation (independent evolution of two branches), where  the input two-parton distributions are generally not known at the low scale $Q_0^2$. For these nonperturbative two-parton functions at low $z_1,z_2$ one may again assume the factorization $D_h^{j_1{'}j_2{'}}(z_1,z_2,Q_0^2) \simeq D_h^{j_1{'}}(z_1,Q_0^2)D_h^{j_2{'}}(z_2,Q_0^2)$
neglecting the constraints due to  momentum conservation ($z_1+z_2<1$). This leads to
\begin{eqnarray} 
\label{DxD_QQ}
& & D_{h1}^{j_1j_2}(x_1,x_2;Q_0^2, Q_1^2, Q_2^2)\nonumber\\
& &~~~ \simeq D^{j_1}_h (x_1; Q_0^2, Q^2_1) D^{j_2}_h (x_2;Q_0^2, Q^2_2).
\end{eqnarray}
\\

As a result of this diagonal (non-skewed) DGLAP evolution of two partons at the DGLAP starting scale $Q_0^2$, the single splitting corrections to the inclusive cross section of TPS read
\begin{widetext}
\begin{eqnarray}
\label{TPSQCD}
& & \sigma^{(A,B,C)}_{\rm TPS,10+01}\nonumber\\
& &= \sum \limits_{i,j,k,l,m,n} \int \sum\limits_{j{'}j_1{'}j_2{'}} \int\limits^{\min(Q_1^2,Q_2^2)}_{Q_0^2} \frac{\alpha_s(k^2)}{2\pi k^2}dk^2 
\int\limits_{x_1}^{1}\frac{dz_1}{z_1}
\int\limits_{x_2}^{1}\frac{dz_2}{z_2} \theta (1-z_1-z_2)
D_h^{j{'}}(z_1+z_2,Q_0^2,k^2)\frac{1}{z_1+z_2} 
P_{j{'} \to j_1{'}j_2{'}}\Bigg(\frac{z_1}{z_1+z_2}\Bigg)\nonumber\\ 
& & \times D_{j_1{'}}^{i}(\frac{x_1}{z_1},k^2,Q_1^2) 
D_{j_2{'}}^{j}(\frac{x_2}{z_2},k^2,Q_2^2) D_{h}^{k}(x_3,Q_0^2,Q_3^2) \hat{\sigma}^A_{il}(x_1, x_1^{'},Q_1^2) 
\hat{\sigma}^B_{jm}(x_2, x_2^{'},Q_2^2) \hat{\sigma}^C_{kn}(x_3, x_3^{'},Q_3^2)\nonumber\\
& & \times D^{l}_{h^{'}} (x_1^{'},Q_0^2, Q_1^2) D^{m}_{h^{'}} (x_2^{'},Q_0^2,Q_2^2) 
D^{n}_{h^{'}} (x_3^{'},Q_0^2,Q_3^2) 
(2\pi)^2 \delta({\bf q_1}+{\bf q_2}+{\bf q_3})\nonumber\\ 
& & \times F_{2g}({\bf q_1}+{\bf q_2}) F_{2g}({\bf q_3}) F_{2g}({\bf -q_1}) F_{2g}({\bf -q_2}) F_{2g}({\bf- q_3}) dx_1 dx_2 dx_3 dx_1^{'} dx_2^{'} dx_3^{'}\frac{d^2q_1}{(2\pi)^2} \frac{d^2q_2}{(2\pi)^2} \frac{d^2q_3}{(2\pi)^2}\nonumber
\end{eqnarray}
\begin{eqnarray}
\label{TPSQCD2}
& & +\sum \limits_{i,j,k,l,m,n} \int \sum\limits_{j{'}j_1{'}j_2{'}} \int\limits^{\min(Q_1^2,Q_3^2)}_{Q_0^2} \frac{\alpha_s(k^2)}{2\pi k^2}dk^2 
\int\limits_{x_1}^{1}\frac{dz_1}{z_1}
\int\limits_{x_3}^{1}\frac{dz_3}{z_3} \theta (1-z_1-z_3)
D_h^{j{'}}(z_1+z_3,Q_0^2,k^2)\frac{1}{z_1+z_3} 
P_{j{'} \to j_1{'}j_2{'}}\Bigg(\frac{z_1}{z_1+z_3}\Bigg)\nonumber\\ 
& & \times D_{j_1{'}}^{i}(\frac{x_1}{z_1},k^2,Q_1^2) 
D_{j_2{'}}^{k}(\frac{x_3}{z_3},k^2,Q_3^2) D_{h}^{j}(x_2,Q_0^2,Q_2^2) \hat{\sigma}^A_{il}(x_1, x_1^{'},Q_1^2) 
\hat{\sigma}^B_{jm}(x_2, x_2^{'},Q_2^2) \hat{\sigma}^C_{kn}(x_3, x_3^{'},Q_3^2)\nonumber\\
& & \times D^{l}_{h^{'}} (x_1^{'},Q_0^2, Q_1^2) D^{m}_{h^{'}} (x_2^{'},Q_0^2,Q_2^2) 
D^{n}_{h^{'}} (x_3^{'},Q_0^2,Q_3^2) (2\pi)^2 \delta({\bf q_1}+{\bf q_2}+{\bf q_3}) \nonumber \\
& & \times F_{2g}({\bf q_1}+{\bf q_3}) F_{2g}({\bf q_2}) F_{2g}({\bf -q_1}) F_{2g}({\bf -q_2}) F_{2g}({\bf- q_3}) dx_1 dx_2 dx_3 dx_1^{'} dx_2^{'} dx_3^{'}\frac{d^2q_1}{(2\pi)^2} \frac{d^2q_2}{(2\pi)^2} \frac{d^2q_3}{(2\pi)^2}\nonumber
\end{eqnarray}
\begin{eqnarray}
\label{TPSQCD3}
& &+ \sum \limits_{i,j,k,l,m,n} \int \sum\limits_{j{'}j_1{'}j_2{'}} \int\limits^{\min(Q_2^2,Q_3^2)}_{Q_0^2} \frac{\alpha_s(k^2)}{2\pi k^2}dk^2 
\int\limits_{x_2}^{1}\frac{dz_2}{z_2}
\int\limits_{x_3}^{1}\frac{dz_3}{z_3} \theta (1-z_2-z_3)
D_h^{j{'}}(z_2+z_3,Q_0^2,k^2)\frac{1}{z_2+z_3} 
P_{j{'} \to j_1{'}j_2{'}}\Bigg(\frac{z_2}{z_2+z_3}\Bigg)\nonumber\\ 
& & \times D_{j_1{'}}^{j}(\frac{x_2}{z_2},k^2,Q_2^2) 
D_{j_2{'}}^{k}(\frac{x_3}{z_3},k^2,Q_3^2) D_{h}^{i}(x_1,Q_0^2,Q_1^2) \hat{\sigma}^A_{il}(x_1, x_1^{'},Q_1^2) 
\hat{\sigma}^B_{jm}(x_2, x_2^{'},Q_2^2) \hat{\sigma}^C_{kn}(x_3, x_3^{'},Q_3^2) \nonumber \\
& & \times D^{l}_{h^{'}} (x_1^{'},Q_0^2, Q_1^2) D^{m}_{h^{'}} (x_2^{'},Q_0^2,Q_2^2) 
D^{n}_{h^{'}} (x_3^{'},Q_0^2,Q_3^2) (2\pi)^2 \delta({\bf q_1}+{\bf q_2}+{\bf q_3})\nonumber \\ 
& & \times F_{2g}({\bf q_1}) F_{2g}({\bf q_2}+{\bf q_3}) F_{2g}({\bf -q_1}) F_{2g}({\bf -q_2}) F_{2g}({\bf- q_3}) dx_1 dx_2 dx_3 dx_1^{'} dx_2^{'} dx_3^{'}\frac{d^2q_1}{(2\pi)^2} \frac{d^2q_2}{(2\pi)^2} \frac{d^2q_3}{(2\pi)^2} \nonumber\\
& & +~...,
\end{eqnarray}
%\end{widetext}
where ``$+...$'' denotes another three analogous contributions, when the splitting takes place for partons (with $x_1^{'}, x_2^{'}, x_3^{'}$) from other colliding hadron $h^{'}$. We do not write them explicitly because of their transparency and to avoid the unnecessary overloading. The integrations over $k^2$ were estimated at the reference  scale $Q_0^2$ (instead of corresponding  ${\bf q_i}^2$) as the lower limit \cite{Ryskin:2011kk,Ryskin:2012qx} due to the strong suppression factors $F_{2g}({\bf q_i})$. 

The second term in Eq.~(\ref{solutiontwoscale}) with one parton at the DGLAP starting scale $Q_0^2$ generates the corrections with two splittings from one hadron %leading to the additional power of the QCD coupling  with respect to Eq.~(\ref{TPSQCD}) 
%and they can be omitted in the first approximation.

\begin{eqnarray}
\label{TPSQCD20}
& & \sigma^{(A,B,C)}_{\rm TPS, 20+02}\nonumber\\
& &= \sum \limits_{i,j,k,l,m,n} \int \sum\limits_{j{'}j_1{'}j_2{'}} \int\limits^{\min(Q_1^2,Q_2^2,Q^2_3)}_{Q_0^2} \frac{\alpha_s(k^2_2)}{2\pi k^2_2}dk^2_2 
\int\limits_{x_1}^{1}\frac{dz_1}{z_1}
\int\limits_{x_2}^{1}\frac{dz_2}{z_2} \theta (1-z_1-z_2)
\sum\limits_{i^{'} i_1 i_3} \int\limits^{k^2_2}_{Q_0^2} \frac{\alpha_s(k^{2}_1)}{2\pi k^{2}_1}dk^{2}_1\nonumber\\
& &\times \int\limits_{z_1+z_2}^{1}\frac{du}{u}\int\limits_{x_3}^{1}\frac{du_3}{u_3} \theta(1-u-u_3) D_h^{i{'}}(u+u_3,Q_0^2,k^{2}_1)\frac{1}{u+u_3} 
P_{i{'} \to i_1 i_3}\Bigg(\frac{u}{u+u_3}\Bigg) D_{i_1}^{j{'}}(\frac{z_1+z_2}{u},k^{2}_1,k^2_2)\nonumber\\
& & \times  \frac{1}{z_1+z_2} 
P_{j{'} \to j_1{'}j_2{'}}\Bigg(\frac{z_1}{z_1+z_2}\Bigg) D_{j_1{'}}^{i}(\frac{x_1}{z_1},k^2_2,Q_1^2) 
D_{j_2{'}}^{j}(\frac{x_2}{z_2},k^2_2,Q_2^2) D_{i_3}^{k}(\frac{x_3}{u_3},k^{2}_1,Q_3^2)\nonumber\\ 
& & \times \hat{\sigma}^A_{il}(x_1, x_1^{'},Q_1^2) 
\hat{\sigma}^B_{jm}(x_2, x_2^{'},Q_2^2) \hat{\sigma}^C_{kn}(x_3, x_3^{'},Q_3^2)\nonumber\\
& & \times D^{l}_{h^{'}} (x_1^{'},Q_0^2, Q_1^2) D^{m}_{h^{'}} (x_2^{'},Q_0^2,Q_2^2) 
D^{n}_{h^{'}} (x_3^{'},Q_0^2,Q_3^2) 
(2\pi)^2 \delta({\bf q_1}+{\bf q_2}+{\bf q_3})\nonumber\\ 
& & \times  F_{2g}({\bf -q_1}) F_{2g}({\bf -q_2}) F_{2g}({\bf- q_3}) dx_1 dx_2 dx_3 dx_1^{'} dx_2^{'} dx_3^{'}\frac{d^2q_1}{(2\pi)^2} \frac{d^2q_2}{(2\pi)^2} \frac{d^2q_3}{(2\pi)^2}~+~...,
\end{eqnarray}

\end{widetext}
where ``$+...$'' denotes again another two contributions with splittings from a hadron $h$ and three contributions with splittings from a hadron $h^{'}$ which can be written by analogy.

One needs to say some words concerning  the expansion in the number of splittings.
When the transverse $q_i$ integral is limited by the form factors, the integral
$\int\alpha_s(k^2)dk^2/k^2$ has indeed the leading logarithm form. However, due to a large anomalous dimension at small $x$ the main contribution comes from a low $k^2$ edge; that is $k^2$ is more close to $Q_0^2$ or to the corresponding $q_i^2$ (recall that in order to keep the leading logarithms in the DGLAP evolution we need $k^2>q^2_i$). By this reason such a contribution may be relatively small. There is also nonperturbative suppression related with the normalization of single distribution functions on the average number of partons since every new additional splitting decreases the number of partons at the the DGLAP starting scale $Q_0^2$.

In addition, there are the corrections with the perturbative QCD splittings from both hadrons (sides). These so-called double splitting contributions can have no strong suppression form factors $F_{2g}({\bf q_i})$ and the integration over ${\bf q_i}$ may be regularized in accordance with our prescription~\cite{Ryskin:2011kk,Ryskin:2012qx} worked out for DPS. 

It makes sense to mention here the discussion~\cite{stir,Blok:2011bu,Gaunt:2012wv,Ryskin:2012qx,Manohar,Gaunt:2012dd,Diehl:2011tt} concerning the double perturbative splitting graphs in the DPS process. Formally, this contribution within the collinear approach in the region of not too small $x$ should be considered as a result of the interaction of {\it one} parton pair with the $2\to 4$ hard subprocess~\cite{stir,Blok:2011bu,Gaunt:2012wv,Ryskin:2012qx,Manohar,Gaunt:2012dd}, since the dominant contribution to the phase space integral comes from a large $q^2\sim \min(Q_1^2,Q_2^2)$. However, due to a larger anomalous dimension generated by the DGLAP evolution at low $x$, actually the major contribution comes from $q^2 << Q^2_1,Q^2_2$ and, as it was argued in Refs.~\cite{Ryskin:2011kk,Ryskin:2012qx}, the contribution under discussion may be validly included in the DPS cross section for appropriately low longitudinal momentum fractions. 

It is worth noticing also that  in the case of TPS the problem of double counting is even more complex in comparison with DPS. We do not concern ourselves with the solution of this problem, and the development of a rigorous theory of TPS here, but are rather interested in understanding the typical structure of scale factors in the case of the two, three and four perturbative QCD splittings. 

For the perturbative QCD splittings from both sides (one from a hadron $h$ and one from a hadron $h'$) we have the nine ($3\times3$) contributions. Their transverse integrations can be easy obtained if one replaces the factorization part $F_{2g}({\bf -q_1}) F_{2g}({\bf -q_2}) F_{2g}({\bf- q_3})$ in the first three terms in Eq.~(\ref{TPSQCD}) with the three single splitting structures 
$F_{2g}({\bf -q_1}-{\bf q_2}) F_{2g}({\bf -q_3})$, $F_{2g}({\bf -q_1}-{\bf q_3}) F_{2g}({\bf -q_2})$ and $F_{2g}({\bf -q_2}-{\bf q_3}) F_{2g}({\bf -q_1})$ consistently. As a result we have the two quite distinctive structure of scale factors: 
\begin{eqnarray} 
\label{fF11diag}
& & \frac{1}{\sigma^2_{\rm TPS, 11, diag}} \\
& &=\int (2\pi)^2 \delta({\bf q_1}+{\bf q_2}+{\bf q_3}) F_{2g}({\bf q_1}+{\bf q_2}) F_{2g}({\bf q_3}) \nonumber \\
& & \times F_{2g}({\bf -q_1}-{\bf q_2})  F_{2g}({\bf- q_3}) \frac{d^2q_1}{(2\pi)^2} \frac{d^2q_2}{(2\pi)^2}\frac{d^2q_3}{(2\pi)^2}\nonumber\\
& & = \int F_{2g}^2({\bf q_3})  F_{2g}^2({\bf- q_3}) \frac{d^2q_1}{(2\pi)^2} \frac{d^2q_3}{(2\pi)^2}\nonumber
\end{eqnarray}
for the three diagonal contributions and
\begin{eqnarray} 
\label{fF11nondiag}
& & \frac{1}{\sigma^2_{\rm TPS, 11, nondiag}} \\
& &=\int (2\pi)^2 \delta({\bf q_1}+{\bf q_2}+{\bf q_3}) F_{2g}({\bf q_1}+{\bf q_3}) F_{2g}({\bf q_2}) \nonumber \\
& & \times F_{2g}({\bf -q_1}-{\bf q_2})  F_{2g}({\bf- q_3}) \frac{d^2q_1}{(2\pi)^2} \frac{d^2q_2}{(2\pi)^2}\frac{d^2q_3}{(2\pi)^2}\nonumber\\
& & = \int  F_{2g}({\bf q_2}) F_{2g}({\bf- q_2}) F_{2g}({\bf q_3})  F_{2g}({\bf- q_3}) \frac{d^2q_2}{(2\pi)^2} \frac{d^2q_3}{(2\pi)^2}\nonumber
\end{eqnarray} 
for the six nondiagonal contributions.

In the case of three diagonal contributions the one transverse integration (${\bf q_1}$ in Eq.~(\ref{fF11diag})) has no strong suppression factors and may be regularized in accordance with our prescription~\cite{Ryskin:2011kk,Ryskin:2012qx} worked out for DPS. 
When we have the diagonal splittings in both sides (that is the ladders form the loop) the integral over the corresponding loop momentum  $q_i$ is limited only by the
condition $q^2_i<k^2$. Neglecting the anomalous dimensions, the $q^2_i$ integral kills the leading logarithms in $k^2$ integrations (on both sides) and finally $k^2$ and $q^2_i$ become of the order of the lowest hard scale $Q^2$ in this ``loop''~\cite{stir,Blok:2011bu,Gaunt:2012wv,Ryskin:2012qx,Manohar,Gaunt:2012dd,Diehl:2011tt}. In other words instead of the DPS (or TPS) we deal with the single parton scattering (SPS) (or DPS) with the $2\to 4$ hard subprocess. However, as was shown in Ref.~\cite{Ryskin:2012qx}, due to a large anomalous dimension at low $x$ relevant for LHC the essential value of $k^2$ may be much smaller than the $Q^2$, leaving the space for the DGLAP evolution between $k^2$ and $Q^2$. To avoid the double counting we have to keep only the contribution with at least one step of DGLAP evolution (one extra $\alpha_s$) between the $k^2$ splitting and $Q^2$ hard $2 \to 2$ subprocess. All this was written in Refs.~\cite{Ryskin:2011kk,Ryskin:2012qx} for the DPS case. However, we recall this discussion once more in order to make the paper complete. Note, that our suggestion is also only appropriate if one has only the most simple ``double splitting'' loop corrections to the SPS. More evolution steps from the DPS are needed to be ``removed'' if one includes higher orders in the SPS. A practical way to do it, i.e. to ``remove'' the appropriate number of overlapping steps of evolution is a subtraction scheme, where a implementation of this appropriate to the DPS/SPS double counting issue has been recently introduced in Ref.~\cite{Diehl:2016khr}.

The six nondiagonal contributions with the perturbative QCD splittings from both sides (like to Eq. (\ref{fF11nondiag})) have strong nonperturbative suppression factors. Nevertheless, they can be also considered as the four-parton interactions. Indeed, here one of the two partons from each hadron $h$ and $h'$ splits into two partons before hard scattering at some scale $k^2$. Then the three hard interactions take place at the different points in the transverse space, but with the small spatial separation $\delta b^2_t \sim 1/k^2$. For comparison in the previous case of diagonal contributions, the two hard interactions take place practically at one point in the transverse space with small spatial separation $\delta b^2_t \sim 1/k^2$, but the third hard interaction takes place at another point with the larger spatial separation $\delta b^2_t \sim 1/m_g^2$ ($m_g$ is the ``effective gluon mass'' ~\cite{Blok:2010ge} which regulates the nonperturbative characteristic scale of the form factors $F_{2g}({\bf q_i})$).

For the three QCD splittings (two from a hadron $h$ and one from a hadron $h'$  and vice versa) we have the eighteen ($ 3 \times 3 \times 2 $) contributions  with the following typical structure of scale factors:
\begin{eqnarray} 
\label{fF21}
& & \frac{1}{\sigma^2_{\rm TPS, 21}} \\
& &=\int (2\pi)^2 \delta({\bf q_1}+{\bf q_2}+{\bf q_3})  \nonumber \\
& & \times F_{2g}({\bf -q_1}-{\bf q_2})  F_{2g}({\bf- q_3}) \frac{d^2q_1}{(2\pi)^2} \frac{d^2q_2}{(2\pi)^2}\frac{d^2q_3}{(2\pi)^2}\nonumber\\
& & = \int F_{2g}({\bf q_3})  F_{2g}({\bf- q_3}) \frac{d^2q_1}{(2\pi)^2} \frac{d^2q_3}{(2\pi)^2}\nonumber.
\end{eqnarray}
Again one can try to regularize the one ``divergent'' transverse integration (${\bf q_1}$ in Eq.~(\ref{fF21})) in accordance with our prescription~\cite{Ryskin:2011kk,Ryskin:2012qx} worked out for DPS. However, in accordance with the discussion~\cite{stir,Blok:2011bu,Gaunt:2012wv,Ryskin:2012qx,Manohar,Gaunt:2012dd,Diehl:2011tt} mentioned above the six ($3 \times 2$) contributions can be also considered as loop corrections to the DPS (to the DPS term with single splitting or to three-parton interactions in the author's terminology of Ref.~\cite{Blok:2013bpa}). This is just  the case where both partons after second splitting from a hadron $h$ interact in hard processes with two partons after splitting from a hadron $h'$ practically at one  point in the transverse space with the small spatial separation and the third parton after first splitting from a hadron $h$ interacts in a hard process with the third parton from a hadron $h'$ at another point in the transverse space and vice versa ($h \to h'$). The other twelve contributions cannot be regarded in such a way as a loop correction to the DPS. They, in fact, overlap with the ``twist 2 $\times$ twist 4'' contributions to the process, which lies one power higher than the TPS. There is also the overlap with some kind of hybrid of ``twist 4 $\times$ twist 2'' and DPS where we have three partons meeting at one point in the transverse space to produce two sets of final state particles in one hard interaction, and two partons meeting at another point in the transverse space to produce the final set of final state particles (which lies at the same power as the TPS). The analysis of DPS on the twist language  can be found in detail, for instance, in Ref.~\cite{Diehl:2016khr}. Of course, the more careful consideration of these contributions demands the special painstaking investigation with not obvious phenomenological issues.

Finally, for the four perturbative QCD splittings both independent transverse integrations have no strong suppression factors at all. The number of four splitting contributions is equal to $3 \times 3 =9$. They overlap with SPS, with three of the contributions also having an overlap with DPS. A full
analysis of these graphs is beyond the scope of the work.

Numerical estimates for the contribution of single perturbative splitting graphs to the DPS cross section were recently done in Ref.~\cite{Blok:2013bpa} (the three-parton interactions in the author's terminology). It was pointed out that the relative contribution of the evolution effects is not small and increases with increasing hard scales and may resolve the long-standing puzzle: why the observed DPS cross section is underestimated by a factor of two in the independent parton approximation (with regard to the ``traditional'' factorization component only)? 

Note that the aim of the paper is not to present some concrete numerical
estimate but to propose a framework for the TPS calculations/description
accounting for the possibility of the $1\to 2$ splitting(s) during the evolution
between the initial scale $Q_0^2$ and the hard scale $Q^2$. As it was shown for the DPS in Ref.~\cite{Ryskin:2012qx} such a splitting may be not negligible at the LHC energies.

Thus, our Eq.~(\ref{TPSQCD}) gives the first estimate for the contribution of the evolution effects to TPS beyond the traditional factorization component and may be useful in further precise investigations. The evolution corrections with the two, three and four perturbative QCD splittings may be taken into account, in principle, also.

\section{\label{sec5} Discussion and conclusions }

The scale factor in Eq.~(\ref{TPSQCD}) in the coordinate representation after the Fourier transformation reads
\begin{eqnarray} 
\label{fFQCD}
& & \frac{1}{\sigma^2_{\rm TPS, 10}} \\
& &=\int (2\pi)^2 \delta({\bf q_1}+{\bf q_2}+{\bf q_3}) F_{2g}({\bf q_1}+{\bf q_2}) F_{2g}({\bf q_3}) \nonumber \\
& & \times F_{2g}({\bf -q_1}) F_{2g}({\bf -q_2}) F_{2g}({\bf- q_3}) \frac{d^2q_1}{(2\pi)^2} \frac{d^2q_2}{(2\pi)^2}\frac{d^2q_3}{(2\pi)^2}\nonumber\\
& &=\int f^2({\bf b -b_3 +b{'}_3})f({\bf b}) f({\bf b_3})f({\bf b{'}_3})d^2b d^2b_3d^2b{'}_3,\nonumber
\end{eqnarray}
and is the same for all six single splitting contributions provided that $f({\bf b_i})$ $(i=1,2,3)$ is supposed to be an universal function for all kind of partons. It should be compared with the scale factor $1/\sigma^2_{\rm TPS, fact}$ for the factorization contribution.

In a simple model where the transverse parton density $f({\bf b_i})$ is taken to have Gaussian functional form
\begin{eqnarray} 
\label{fgauss}
& & f({\bf b_i})= \exp{(-b_i^2/R^2)}/\pi R^2,i=1,2,3,\nonumber\\
& & F_{2g}({\bf q_i})=\exp{(-q_i^2R^2/4)},
\end{eqnarray}
all scale factors are calculated analytically:
\begin{eqnarray} 
\label{scale}
\frac{1}{\sigma^2_{\rm TPS, 10}}=\frac{1}{7\pi^2R^4}, ~~\frac{1}{\sigma^2_{\rm TPS, fact. 00}}=\frac{1}{12\pi^2R^4}=\frac{4}{3\sigma^2_{\rm eff}}.
\end{eqnarray}
Their ratio
\begin{eqnarray} 
\label{scaleR}
\frac{\sigma^2_{\rm TPS, fact. 00}}{\sigma^2_{\rm TPS, 10}}=\frac{12}{7}=1.7
\end{eqnarray}
shows that the single splitting contributions to the cross section are enhanced, relative to the factorization one, by the factor
\begin{eqnarray} 
\label{enhan}
2 \times 3 ({\rm combinatorial}) \times 1.7 ({\rm scale}) \sim 10.
\end{eqnarray}
Let us remind that in the case of DPS the similar enhancement factor is equal to
\begin{eqnarray} 
\label{enhanDPS}
2 ({\rm combinatorial}) \times 2 ({\rm scale}) = 4
\end{eqnarray}
and is not strongly sensitive~\cite{Blok:2011bu,Gaunt:2012dd,Gaunt:2014rua} to the functional form of the transverse parton density.

So we conclude that the single splitting terms may provide a sizable contribution to the cross section of TPS even if they constitute a small correction to the triple factorized parton distribution functions.

Other nonperturbative scale factors incoming in Eqs.~(\ref{TPSQCD20}) and (\ref{fF11nondiag}) are  easily calculated also in a simple model with the Gaussian transverse parton density. They are displayed in Table I together with the appropriate combinatorial factors for easy comparison.
%\begin{eqnarray} 
%\label{scale2}
%\frac{1}{\sigma^2_{\rm TPS, 20}}=\frac{1}{3\pi^2R^4}, ~~\frac{1}{\sigma^2_{\rm TPS, 11, %nondiag}}=\frac{1}{4\pi^2R^4}.
%\end{eqnarray}

\begin{table}
\caption{\label{tab:table1}
The scale and combinatorial factors for different contributions under consideration.
 }
\begin{ruledtabular}
\begin{tabular}{ccccc}
        & Scale factor $\times \pi^2 R^4$ &  Comb. factor  \\
 %\tableline
\colrule
fact. (00) contr.        & 1/12     & 1     \\
(10+01) contr.                  & 1/7     & 6     \\
(20+02) contr.        & 1/3     &    6  \\
(11), nondiag. contr. & 1/4    &  6      \\
\end{tabular}
\end{ruledtabular}
\end{table}

In summary, we suggest a practical method which makes it possible to estimate the inclusive cross section for TPS, taking into account the QCD evolution and based on the well-known collinear distributions. Numerical estimations~\cite{Ryskin:2012qx,Blok:2013bpa,Gaunt:2014rua,sjostrand3} of these evolution effects, done in the case of DPS in the single splitting approximation, support strongly the significance of such calculations for TPS also at the LHC energy, for instance, for the final states containing 6 jets or 3 $c {\bar c}$ pairs.

\begin{acknowledgments}
It is a pleasure to thank M.G.~Ryskin for careful reading of the manuscript, valuable comments and corrections. Discussions with A.I.~Demianov, A.P.~Kryukov and G.M.~Zinovjev are gratefully acknowledged. It is also a pleasure to thank  the referee for deep and interesting suggestions and questions.
\end{acknowledgments}

%\clearpage


\begin{thebibliography}{99}
\bibitem{Bartalini:2011jp} P. Bartalini {\it et al.,} arXiv:1111.0469 [hep-ph].
\bibitem {Abramowicz:2013iva} H.~Abramowicz {\it et al.,}  arXiv:1306.5413 [hep-ph].
\bibitem{Bansal:2014paa} S.~Bansal {\it et al.,} arXiv:1410.6664 [hep-ph].
\bibitem{Astalos:2015ivw} R~Astalos {\it et al.,} arXiv:1506.05829 [hep-ph].
\bibitem{AFS} T.~Akesson {\it et al.} (AFS Collaboration), Z. Phys. C {\bf 34}, 163 (1987).
\bibitem{UA2} J.~Alitti {\it et al.} (UA2 Collaboration),  Phys. Lett. B {\bf 268}, 145 (1991).
\bibitem{cdf4jets} F.~Abe {\it et al.} (CDF Collaboration), Phys. Rev. D {\bf 47}, 4857 (1993).
\bibitem{cdf} F.~Abe {\it et al.} (CDF Collaboration), Phys. Rev. D {\bf 56}, 3811 (1997).
\bibitem{D0} V.M.~Abazov {\it et al.} (D0 Collaboration), Phys. Rev. D {\bf 81},
052012 (2010).
\bibitem{D01} V.M.~Abazov {\it et al.} (D0 Collaboration), Phys. Rev. D {\bf 83}, 052008 (2011).
\bibitem{D02} V.M.~Abazov {\it et al.} (D0 Collaboration), Phys. Rev. D {\bf 89}, 072006 (2014)
\bibitem{atlas} G.~Aad {\it et al.} (ATLAS Collaboration), New J. Phys. {\bf 15}, 033038 (2013).
\bibitem{cms} S.~Chatrchyan {\it et al.} (CMS Collaboration), J. High Energy Phys. 03 (2014) 032.
\bibitem{Abelev}
B.~Abelev {\it et al.} (ALICE Collaboration), Phys. Lett. B {\bf 712}, 165 (2012).
\bibitem{Aaij} R.~ Aaij {\it et al.} (LHCb Collaboration), Phys. Lett. B {\bf 707}, 52 (2012).
\bibitem{kom} C.-H.~Kom, A.~Kulesza, and W.J.~Stirling, Phys. Rev. Lett. {\bf 107}, 082002 (2011).
\bibitem{Baranov:2011ch} S.P.~Baranov, A.M.~Snigirev, and N.P.~Zotov, Phys. Lett. B {\bf 705}, 116 (2011).
\bibitem{Novoselov} A.A.~Novoselov, arXiv:1106.2184 [hep-ph].
\bibitem{Baranov:2012re} S.P.~Baranov, A.M.~Snigirev, N.P.~Zotov, A.~Szczurek, and W.~Schafer, Phys. Rev. D {\bf 87}, 034035 (2013).
\bibitem{Calucci:2009sv} G.~Calucci and D.~Treleani, Phys. Rev. D {\bf 79}, 074013 (2009).
\bibitem{Calucci:2009ea} G.~Calucci and D.~Treleani, Phys. Rev. D {\bf 80}, 054025 (2009).
\bibitem{Treleani:2012zi} D.~Treleani and G.~Calucci, Phys. Rev. D {\bf 86}, 036003 (2012).
\bibitem{Maina:2009sj} E.~Maina, J. High Energy Phys. 09  (2009) 081.
\bibitem{Ryskin:2011kk} M.G.~Ryskin and A.M.~Snigirev, Phys. Rev. D {\bf 83}, 114047 (2011).
\bibitem{Ryskin:2012qx} M.G.~Ryskin and A.M.~Snigirev, Phys. Rev. D {\bf 86}, 014018 (2012).
\bibitem{Blok:2010ge}B.~Blok, Yu.~Dokshitzer, L.~Frankfurt, and M.~Strikman, Phys. Rev. D {\bf 83}, 071501 (2011).
\bibitem{Kirschner:1979im} R.~Kirschner, Phys.\ Lett.\  {\bf 84B}, 266 (1979).
\bibitem{Shelest:1982dg} V.P.~Shelest, A.M.~Snigirev, and G.M.~Zinovjev, Phys.\ Lett.\  {\bf 113B}, 325 (1982); Theor. Math. Phys. {\bf 51}, 523 (1982) [Teor. Mat. Fiz.{\bf 51}, 317 (1982)].
\bibitem{gribov} V.N.~Gribov and L.N.~Lipatov, Sov. J. Nucl. Phys. {\bf 15}, 438 (1972) [Yad. Fiz. {\bf 15}, 781 (1972)]; {\bf 15}, 675 (1972) [Yad. Fiz. {\bf 15}, 1218 (1972)].
\bibitem{lipatov} L.N.~Lipatov, Sov. J. Nucl. Phys. {\bf 20}, 94 (1974) [Yad. Fiz. {\bf 20}, 181 (1974)].
\bibitem{dokshitzer} Yu.L.~Dokshitzer, Sov. Phys. JETP {\bf 46}, 641 (1977) [Zh. Eksp. Teor. Fiz. {\bf 73}, 1216 (1977)].
\bibitem{altarelli} G.~Altarelli and G.~Parisi, Nucl. Phys. {\bf B126}, 298 (1977) 
\bibitem{snig2} V.P.~Shelest, A.M.~Snigirev, and G.M.~Zinovjev,  Report No. ITP-83-46E, Kiev, 1983 (unpublished). 
\bibitem{Blok:2011bu} B.~Blok, Yu.~Dokshitzer, L.~Frankfurt, and M.~Strikman, Eur. Phys. J. C {\bf 72}, 1963 (2012).
\bibitem{Gaunt:2012wv} J.R.~Gaunt and W.J.~Stirling, arXiv:1202.3056 [hep-ph].
\bibitem{Gaunt:2012dd} J.R.~Gaunt, J. High Energy Phys. 01 (2013) 042.
\bibitem{stir} J.R.~Gaunt and W.J.~Stirling, J. High Energy Phys. 06 (2011) 048.
\bibitem{Manohar} A.V.~Manohar and W.J.~Waalewijn, Phys. Lett. B {\bf 713}, 196 (2012).
\bibitem{Diehl:2011tt} M.~Diehl and A.~Schafer, Phys. Lett. B {\bf 698}, 389 (2011).
\bibitem{Diehl:2016khr} M.~Diehl and J.R.~Gaunt, arXiv:1603.05468 [hep-ph].
\bibitem{Blok:2013bpa}B.~Blok, Yu.~Dokshitzer, L.~Frankfurt, and M.~Strikman, Eur. Phys. J. C {\bf 74}, 2926 (2014).
%bibitem{Snigirev:2010ds} A.M.~Snigirev, Phys. Rev. D {\bf 83}, 034028 (2011).
\bibitem{Gaunt:2014rua} J.R.~Gaunt, R.~Maciula and A.~Szczurek, Phys. Rev. D {\bf 90}, 054017 (2014).
\bibitem{sjostrand3} T.~Sjostrand and P.Z.~Skands, Eur. Phys. J. {\bf C 39}, 129 (2005).
\end{thebibliography}
\end{document}